\documentclass[12pt]{iopart}
\usepackage{graphicx}
\usepackage{iopams}

\begin{document}

\title[Circles of equal radii randomly placed on a plane]{Circles of equal radii randomly placed on a plane: some rigorous results, asymptotic behavior, and application to transparent electrodes}

\author{R K Akhunzhanov, Y Y Tarasevich, I V Vodolazskaya}

\address{Laboratory of  Mathematical Modeling, Astrakhan State University, Astrakhan, 414056, Russia}
\ead{tarasevich@asu.edu.ru}
\vspace{10pt}
\begin{indented}
\item[]July 2019
\end{indented}

\begin{abstract}
We consider $N$ circles of equal radii, $r$, having their centers randomly placed within a square domain $\mathcal{D}$ of size $L \times L$ with periodic boundary conditions ($\mathcal{D} \in \mathbb{R}^2$). When two or more circles intersect each other, each circle is divided by the intersection points into several arcs. We found the exact length distribution of the arcs. In the limiting case of dense systems and large size of the domain  $\mathcal{D}$ ($L \to \infty$ in such a way that the number of circle per unit area, $n=N/L^2$, is constant), the arc distribution approaches the probability density function (PDF) $f(\psi) = 4 n r^2\exp(-4 n r^2 \psi)$, where $\psi$ is the central angle subtended by the arc. This PDF is then used to estimate the sheet resistance of transparent electrodes based on conductive rings randomly placed onto a transparent insulating film.
\end{abstract}

%
\vspace{2pc}

\submitto{}
%
%

\section{Introduction}\label{sec:intro}

Problems relating to the covering of circles by randomly placed arcs of random length have been  solved over recent  decades~\cite{Dvoretzky1956PNAS,Shepp1972IJM,Flatto1973IJM,Solomon1978,Holst1980JAProb,Siegel1982JAProb,Hall1984SPTA,Holst1984JSM,Holst1984JAProb,Huffer1987JAProb,Huillet2003JPhysA,Huillet2003AAP,Durand2010}.
Despite the seeming academicism, such problems have numerous applications. Thus, in~\cite{Yadin1982JAProb}, covering a circle by randomly placed arcs was considered in connection with its application to the shading problem; in~\cite{Hall1984SPTA,Huillet2003JPhysA}, the problem was associated with a random sequential adsorption of rods and a parking problem; in~\cite{Moriarty2007}, the approach was applied to the problems of genomics. Related problems of covering surfaces with circles are directly related to networks of sensors~\cite{Wang2011ACMCS} and mobile connections~\cite{Jhuang2017IEEE}.

Films constituting a conductive mesh on a transparent insulating substrate are known as transparent electrodes. The mesh can be produced by different methods such as ink-jet printing, deposition, etc. A particular case of such electrodes is that of ring-based electrodes. Conductive rings can be created, e.g., by using the ``coffee-ring effect''~\cite{Layani2009ACSN,Shimoni2014} or by chemical methods~\cite{Azani2018ChEJ,Azani2019JAP}.

The rest of this paper is constructed as follows. In~\sref{sec:methods}, the model is described and all necessary quantities are defined. In~\sref{sec:results}, we consider the intersection of only two circles (\sref{subsec:circle2}); then, we obtain results for a circle that is intersected by $k$ other circles (\sref{subsec:circles-k}); finally, we derive a rigorous formula for a system of $N$ circles (\sref{subsec:circles-N}). Asymptotic behavior is studied in \sref{subsec:asymptotic}. \Sref{sec:application} is devoted to the sheet resistance of ring-based transparent electrodes. \Sref{sec:conclusion} summarizes our main results.

\section{Model, definitions, and methods}\label{sec:methods}

Consider a square domain $\mathcal{D}$ of a plane $\mathbb{R}^2$. Let the size of the domain be  $L\times L$. The domain is subject to periodic boundary conditions (PBCs). PBCs are applied to simplify the consideration by eliminating any effects of borders.

The centers of the $N$ circles of radii $r$ are assumed to be independent and identically distributed (i.i.d.) in $\mathcal{D}$, i.e., $x,y \in [0;L]$, where $(x,y)$ are the coordinates of the center of the circle. The relation $L>2r$ is assumed.

The number of circles per unit area
\begin{equation}\label{eq:n}
  n = \frac{N}{L^2}
\end{equation}
is known as the number density. When asymptotic behavior is considered, any changes of the quantities $L$ and $N$ are supposed to be consistent in such a way as ensure preservation of the given value of the number density,~$n$.

If some circles intersect each other, these circles are divided into several arcs. An isolated circle is supposed to have only one arc of length $2\pi r$. We are looking for the length distribution of these arcs, and can characterize any arc by the angle that it subtends at the center of the circle. All angles are measured counterclockwise. For simplicity, when we refer to the length of an arc, we mean the angular distance of that arc. Since we are dealing only with circular arcs, we shall simply use the term ``arc''.

An outline of our ``road map'' is as follows. To obtain the length distribution of the arcs in a system of $N$ circles, we are going to find the length distribution of the arcs in only one circle that intersects exactly $k$ other circles. With such a distribution in hand, we can apply a binomial distribution to obtain the arc length distribution in a system of $N$ circles. We will start with the simplest case $k=1$, then, we will add some extra circles.

\section{Rigorous results}\label{sec:results}
\subsection{Two intersecting circles}\label{subsec:circle2}
\begin{figure}[!hb]
  \centering
  \includegraphics[width=\textwidth]{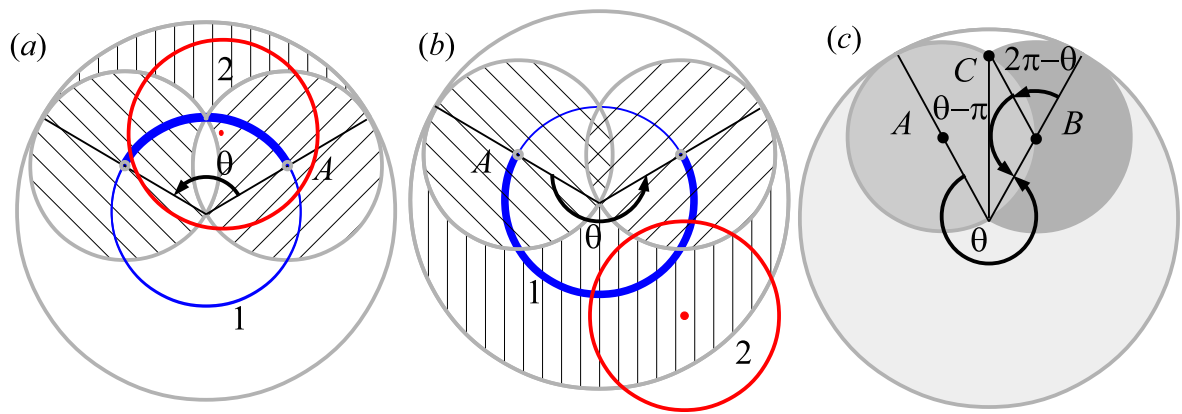}
  \caption{Diagram to indicate the calculation of the probability that the minimal angle between an arbitrary point $A$ on circle 1 and the intersection point with circle 2 is less than $\theta$. $(a)$ $0\leqslant\theta\leqslant \pi$; $(b)$ $\pi<\theta\leqslant 2\pi$; $(c)$ shows some angles under consideration.}\label{fig:angles}
\end{figure}

Let a circle (circle~1) be intersected by another circle (circle~2). Let us consider an arbitrary  point $A$ on circle~1. Let an arc $\theta$ starts from $A$. When  $0 \leqslant \theta \leqslant \pi$, circle~2 intersects or touches circle~1  within this arc if and only if (iff) the center of the circle~2 is located within the hatched region in \fref{fig:angles}$(a)$. The area of the hatched region is the area of the sector with the central angle $\theta$ ($2\theta r^2$) plus the area of two semicircles with radii $r$, i.e. the area of the circle with radius $r$ ($\pi r^2$), minus the area of the region in the form of a \emph{vesica piscis} ($r^2 (\pi - \theta -\sin\theta)$). Hence, the probability that the intersection point is located between 0 and $\theta$ from point $A$ is
$$
  F = \frac{3\theta + \sin\theta}{4 \pi}.
$$

When $\pi<\theta\leqslant 2\pi$ (\fref{fig:angles}$(b)$), circle~2 intersects or touches circle~1 within the arc $\theta$ iff  the center of circle~2 locates within the hatched region in~\fref{fig:angles}$(b)$. The area of this region is equal to the area of the sector $\theta$ ($2r^2 \theta$), plus the area of two isosceles triangles ($r^2 \sin(\theta - \pi)$), plus the area of the two sectors ($r^2 (2\pi - \theta )$) (see \fref{fig:angles}$(c)$). Hence, the probability that the intersection point is located between 0 and $\theta$ from point $A$ is
$$
  F = \frac{2\pi + \theta - \sin \theta}{4 \pi}.
$$
Finally,
\begin{equation}\label{eq:Ftheta1}
F_\Theta(\theta;1) = \frac{1}{4\pi}
\cases{3\theta + \sin(\theta), & $0\leqslant\theta\leqslant \pi;$\\
2\pi + \theta - \sin(\theta), & $\pi<\theta\leqslant 2\pi.$}
\end{equation}
the PDF can be easily obtained  from~\eref{eq:Ftheta1} by differentiation
\begin{equation}\label{eq:ftheta1}
f_\Theta(\theta;k=1) = \frac{1}{4\pi}
\cases{
 3 + \cos\theta,\\
1 - \cos\theta.
}
\end{equation}
Hereinafter, we suppose that the upper line corresponds to the case $0\leqslant\theta\leqslant \pi$, while the bottom line corresponds to the case $\pi<\theta\leqslant 2\pi$.

\subsection{Given circle is intersected by $k$ other circles}\label{subsec:circles-k}
Let there be a circle (circle~1) that is intersected by exactly $k$ other circles ($1 \leqslant k  \leqslant N-1 $). Let $A$ be an arbitrary point on circle~1 and $\Theta_k$ be the angle between this point and the nearest (counterclockwise) intersection point with another circle ($0\leqslant\Theta_k\leqslant 2\pi$). In fact, the case $k=1$ was considered in~\sref{subsec:circle2}, hence, now we start from $k=2$.

\begin{figure}[!hb]
  \centering
  \includegraphics[width=\textwidth]{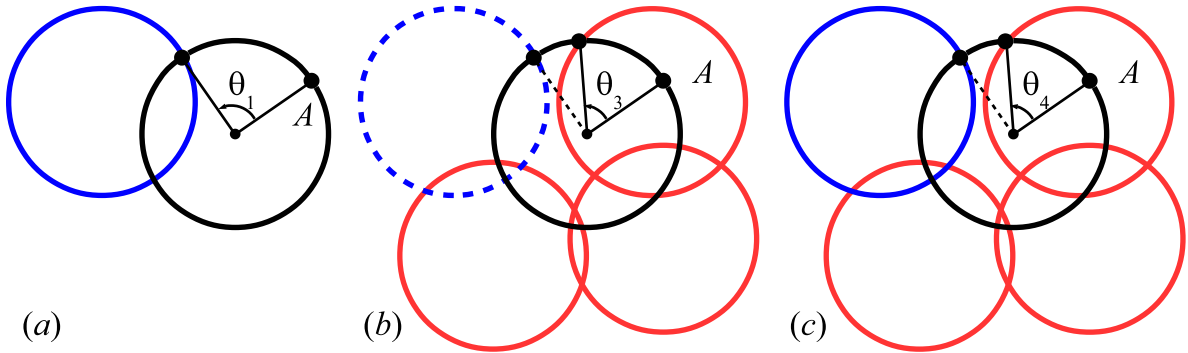}
  \caption{Angle between given point $A$ on circle~1 and the nearest intersection point, when  circle~1 is intersected by exactly $k$ other circles. $(a)$~$k=1$; $(b)$~$k=3$; $(c)$~$k=4$.}\label{fig:intersections}
\end{figure}

Let us divide $k$ circles into two groups, viz., the first group consists of only one circle, while the other group consists of the rest of the $k-1$ circles. For each of the two groups, $\Theta_1$ and $\Theta_{k-1}$ are the angles between point $A$ and the nearest intersection point, respectively. $\Theta_1$ and $\Theta_{k-1}$ are assumed to be independent. $\Theta_k =\min(\Theta_1,\Theta_{k-1})$ (see~\fref{fig:intersections}).

If $A$, $B$, and $C$ are random variables with the cumulative distribution functions (CDFs) $F_A, F_B, F_C$ and $A = \max (B,C)$, then $F_A(x) = F_B(x) F_C(x)$. Respectively,
$1 - F_A(x) = (1 - F_B(x)) ( 1 - F_C(x))$, if $A = \min (B,C)$.
Hence,
$$
1 - F_\Theta(\theta; k)=(1 - F_\Theta(\theta;1))(1 - F_\Theta(\theta;k-1)).
$$
Thereby
$$
1 - F_\Theta(\theta;k)=(1 - F_\Theta(\theta;1))^k.
$$
Accounting~\eref{eq:Ftheta1}, the CDF is
\begin{equation}\label{eq:FTheta}
F_\Theta(\theta;k) = 1- (1 - F_\Theta(\theta;1))^k =
1 - \frac{1}{(4\pi)^k} \cases{
\left(4\pi - 3\theta - \sin\theta\right)^k,\\
\left(2\pi - \theta + \sin\theta\right)^k.
}
\end{equation}
Hence, the PDF is
\begin{equation}\label{eq:rhoTheta}
f_\Theta(\theta;k) =
\frac{k}{(4\pi)^k} \cases{
\left(4\pi - 3\theta - \sin\theta\right)^{k-1} (3+\cos\theta),\\
\left(2\pi - \theta + \sin\theta\right)^{k-1} (1-\cos\theta).
}
\end{equation}

\subsection{System of $N$ circles}\label{subsec:circles-N}
\begin{figure}[!hb]
  \centering
  \includegraphics[width=0.5\textwidth]{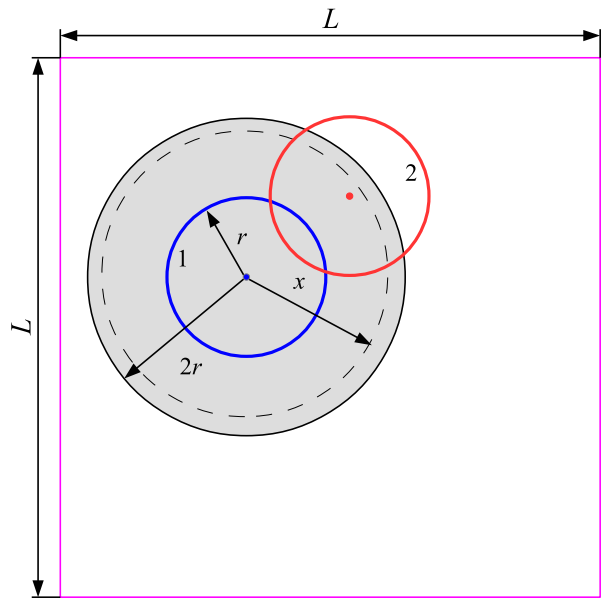}
  \caption{Two circles intersect each other, iff the distance between their centers $l \leqslant 2r$. If circle 1 and circle 2 intersect each other, the probability that the distance between their centers does not exceed $x$ is the ratio of the area enclosed by a circle of radius $x$  to the area enclosed by a circle of radius $2r$.}\label{fig:ring}
\end{figure}

If circle 1 and circle 2 intersect each other (\fref{fig:ring}), the probability that the distance, $l$, between their centers does not exceed $x$ is
$$
\Pr (l \leqslant x)  = \frac{x^2}{4 r^2}.
$$
Hence, the distance, $x$, between the centers of the two intersecting circles obeys the PDF
$$
f_X(x)=\frac{x}{2r^2},
$$
where $0\leqslant x\leqslant 2r$, while the PDF of the central angle is
\begin{equation}\label{eq:rhosin}
  f_\Phi(\varphi) = \frac{\sin\varphi}{2}, \quad 0\leqslant\varphi\leqslant\pi
\end{equation}
(see~\fref{fig:ring}).

Let the random variable $\Psi$ be the central angle corresponding to a random arc, $0\leqslant\Psi\leqslant 2\pi.$ Let a given circle (circle~1) be intersected by exactly $k$ other circles. The intersection points between circle~1 and these $k$ circles divide circle~1 into $2k$ arcs. The random variable  $\Psi_k$ is the central angle of a random arc produced by the intersection of an arbitrary circle with exactly  $k$ other circles. Its PDF  is $f_{\Psi}(\psi;k)$, where $0 \leqslant k  \leqslant N-1 $. The PDF of an isolated circle (with central angle of the arc $2\pi$ by agreement) is the  $\delta$-function
$$
f_\Psi(\psi;0) = \delta(\psi - 2\pi).
$$

Let circle~1 be intersected by only one other circle. In this case, circle~1 is divided into two arcs, viz.,  $\psi$ and $2\pi - \psi$. Accounting~\eref{eq:rhosin}, the PDF is
\begin{equation}\label{eq:rhoPsi}
f_\Psi(\psi;1) =\frac{1}{2}
\cases{
f_\Phi(\psi),\\
f_\Phi(2\pi - \psi),
}
 =\frac{\sin\psi}{4}
\cases{
1,\\
-1.
}
\end{equation}
The CDF can be found by integrating~\eref{eq:rhoPsi}
\begin{equation}\label{eq:FPsi1}
F_\Psi(\psi;1) =
\frac{1}{4}\cases{
1-\cos\psi,\\
3+\cos\psi.
}
\end{equation}

\begin{figure}[!htb]
  \centering
  \includegraphics[width=0.6\textwidth]{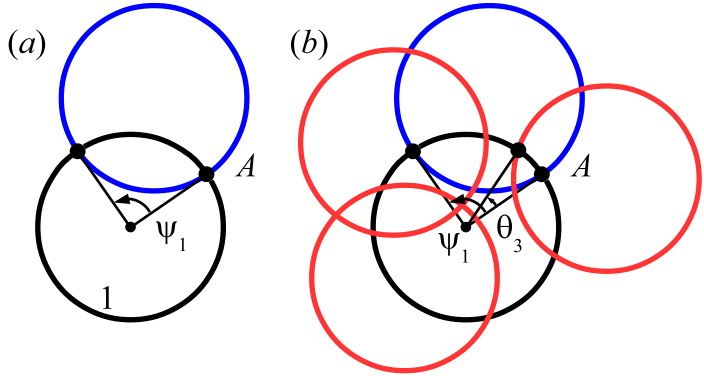}
  \caption{$(a)$ Central angle $\psi_1$  in circle~1 between two intersection points with other circle. $(b)$ Central angle $\theta_3$  when circle~1 is intersected by exactly $k$ other circles ($k=3$).  $\Psi_4 = \min(\Psi_1, \Theta_3)$.}\label{fig:intersections2}
\end{figure}

Let us consider a circle (circle 1). Add one circle to intersect circle~1. Choose one of the two intersection points as a given point (point $A$). Now, add an extra $k-1$ circles intersecting circle~1 (\fref{fig:intersections2}), then
$$
\Psi_k = \min(\Psi_1, \Theta_{k-1}).
$$
Here, $2 \leqslant k  \leqslant N-1 $.
Both the random variables $\Psi_1$ and $\Theta_{k-1}$ are assumed to be independent. Then,
\begin{equation}\label{eq:1F}
1 - F_\Psi(\psi;k)=(1 - F_\Psi(\psi;1))(1 - F_\Theta(\theta; k - 1)).
\end{equation}

When $k \geqslant 2$, the PDF can be obtained using differentiation of the CDF~\eref{eq:1F}
$$
f_\Psi(\psi;k) = (1 - F_\Theta(\psi;k-1))f_\Psi(\psi;1) +
(1 - F_\Psi(\psi;1))f_\Theta(\psi;k-1).
$$
Accounting~\eref{eq:FTheta}, \eref{eq:rhoTheta},\eref{eq:rhoPsi}, and~\eref{eq:FPsi1},
\begin{equation}\label{eq:rhoPsim}
\eqalign{
\fl  f_\Psi(\psi;k) =
\frac{\sin\psi}{4({4\pi})^{k-1}}\cases{
( 4\pi - 3\psi - \sin\psi )^{k-1}\\
-( 2\pi - \psi + \sin\psi )^{k-1}
}+ \\
\frac{k-1}{4({4\pi})^{k-1}}
\cases{
( 3+\cos\psi)^2 ( 4\pi - 3\psi - \sin\psi )^{k-2} \\
( 1-\cos\psi)^2 ( 2\pi - \psi + \sin\psi )^{k-2}.
}
}
\end{equation}
It is noteworthy that, despite the assumption $k \geqslant 2$, \eref{eq:rhoPsim} is also correct for $k=1$ transforming into~\eref{eq:rhoPsi}.

Two circles intersect each other, iff the distance between their centers $l \leqslant 2r$, i.e., the center of the second circle is located inside a circle of radius $2r$ concentric with the first circle (the shadowed region in \fref{fig:ring}). Since the area of the shadowed region is $4\pi r^2$, the probability of intersection of two arbitrary circles randomly located in  $\mathcal{D}$ is
\begin{equation}\label{eq:P}
  P = 4\pi \left( \frac{r}{L} \right)^2.
\end{equation}
Accordingly, the probability that the two circles do not intersect each other is equal to
\begin{equation}\label{eq:Q}
Q = 1 - P.
\end{equation}
The probability, $P_k$, that a given circle is intersected by exactly $k$ other circles follows the binomial distribution
\begin{equation}\label{eq:Pk}
P_k = {{N-1}\choose k}  P^k Q^{N-1-k}.
\end{equation}

Let circle~1 be intersected by exactly $k$ other circles. Then, circle~1 is divided into $\nu(k)$ arcs
$$
\nu(k) = 2k + \delta_{k0},
$$
where $ \delta_{ij}$ is  the Kronecker delta.

The random variable $M$ is the number of circles intersecting a random arc, $0\leqslant M\leqslant N-1$. Choose a random arc and consider the circle that this arc belongs to. Find the probability that this circle is intersected by $m$ other circles. The expected total number of arcs in the system is
$$
\sum_{k=0}^{N-1}\nu(k)P_k.
$$
The expected total number of arcs belonging to the arcs divided exactly into $\nu(m)$ arcs
is $\nu(m)P_m$. The probability of interest is
$$
\Pr(M = m) = \frac{\nu(m)P_m }{\sum_{k=0}^{N-1}\nu(k)P_k}
= \frac{ P_0 \delta_{m0} + 2 m P_m }{P_0 + 2\sum_{k=1}^{N-1}k P_k}.
$$

The PDF, that a random arc subtending an angle $\psi$ upon the condition that the arc belongs to a circle intersected by exactly $m$ other circles, is equal to $f_{\Psi}( \psi; m)$.
Application of the formula of total probability gives
\begin{equation*}
\eqalign{
f_{\Psi}(\psi)
 = \sum_{m=0}^{N-1} f_{\Psi}( \psi; m)  \Pr(M = m) = \\
 = \sum _{m=0}^{N-1} f_{\Psi}( \psi; m) \frac{P_0 \delta_{m0} + 2 m P_m }{P_0 + 2\sum_{k=1}^{N-1}k P_k}=
  \frac{ P_0 f_{\Psi}( \psi; 0) + 2 \sum  _{m=1}^{N-1} f_{\Psi}( \psi; m) m P_m }{P_0 + 2\sum_{k=1}^{N-1}k P_k}.
  }
\end{equation*}
Using obtained formulae for $P_k$ \eref{eq:Pk} and $f_{\Psi}(\psi;k)$~\eref{eq:rhoPsim}, this transforms into
\begin{equation}\label{eq:fpsi}
\fl
\eqalign{
f_{\Psi}(\psi)
=\frac{1}{Q^{N-1} + 2(N-1)P} \times\\\left(  Q^{N-1}  \delta(\psi - 2\pi) +
\frac{
(N-1) P \sin\psi}{2}
\cases{
\left(1 - \frac{3\psi + \sin\psi}{4\pi}P \right) ^{N-2} \\
- \left(1 + \frac{-2\pi - \psi + \sin\psi}{4\pi}P \right) ^{N-2}
} +\right.\\
+\left.\frac{ (N-1)(N-2) P^2 }{8\pi}
\cases{
\left( 3+\cos\psi \right)^2
\left( 1 - \frac{3\psi + \sin\psi}{4\pi}P \right) ^{N-3} \\
\left( 1-\cos\psi\right)^2
\left( 1 + \frac{-2\pi - \psi + \sin\psi}{4\pi}P \right) ^{N-3}
}
\right).
}
\end{equation}
\Eref{eq:fpsi} together with definitions~\eref{eq:P} and~\eref{eq:Q} is the PDF of the central angles of the arcs produced by the intersections of $N$ circles of equal radii $r$ with i.i.d. centers within $\mathcal{D}$.

\subsection{Asymptotic behavior}\label{subsec:asymptotic}

Let $L$, the size of the domain under consideration $\mathcal{D}$, tends to infinity in such a way that $n = \mathrm{const}$, i.e., the number of circles varies as $N = n L^2$. Application of the $\lim_{x \to \infty} \left(1+ x^{-1}\right)^x = \e$ to~\eref{eq:fpsi} leads to the PDF  ($\lim_{L\to\infty}f_{\Psi}(\psi)= f^\infty_{\Psi}(\psi,n)$)
\begin{equation}\label{eq:fPsiLinfty}
\fl
\eqalign{
f^\infty_{\Psi}(\psi,n)
=\frac{1}{1 + 8\pi r^2 n \exp(4\pi r^2 n)}\delta(\psi - 2\pi) + \\
\frac{ 2\pi r^2 n}{\exp(-4\pi r^2 n) + 8\pi r^2 n}
\cases{
(\sin\psi + (3+\cos\psi)^2 r^2 n)  \exp(- r^2 n (3\psi + \sin\psi)), \\
(-\sin\psi + ( 1-\cos\psi)^2 r^2 n )  \exp(- r^2 n (2\pi + \psi - \sin\psi)).
}
}
\end{equation}

\begin{figure}[!hbtp]
  \centering
  \includegraphics[width=0.6\textwidth]{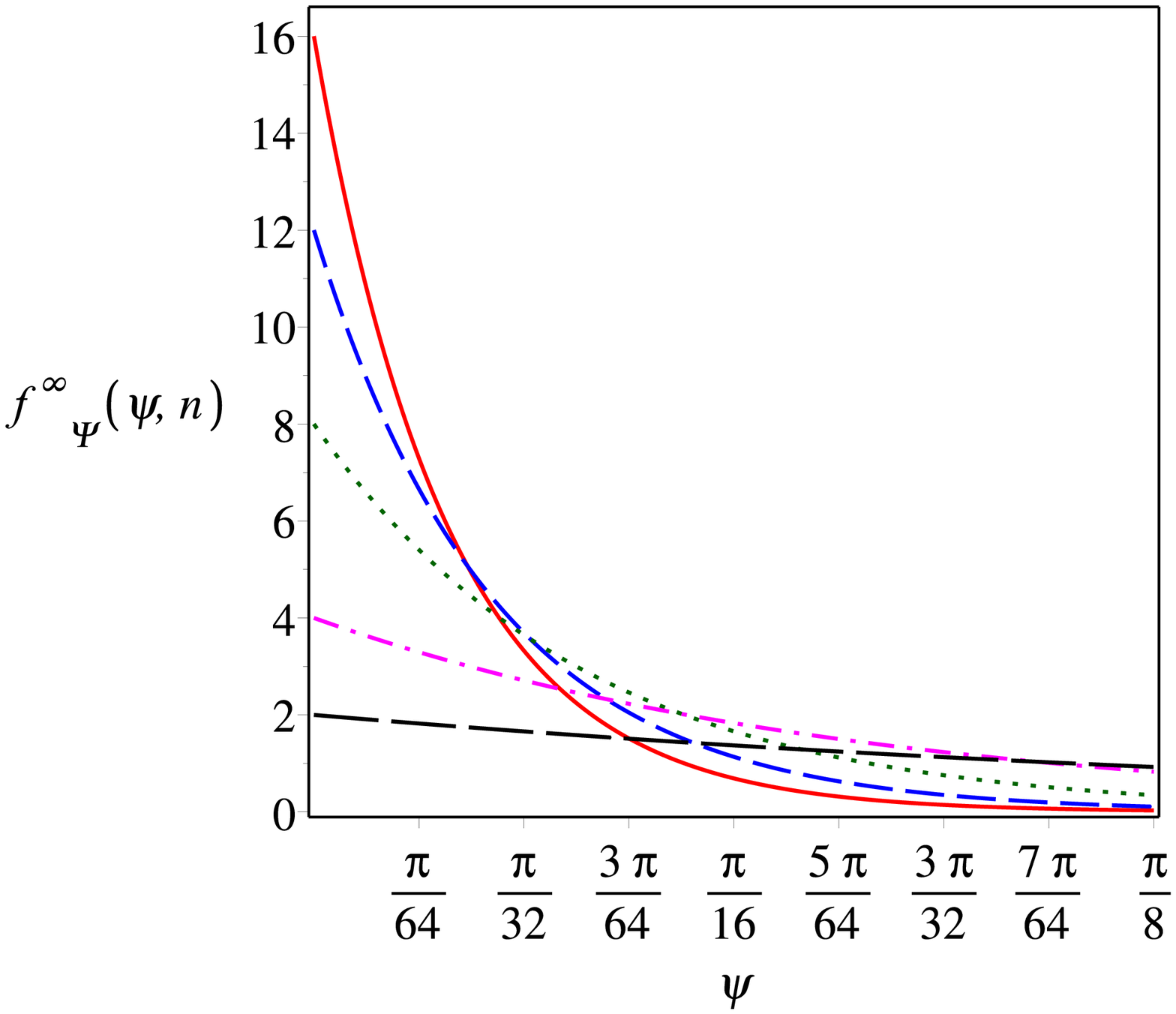}
  \caption{PDFs~\eref{eq:fPsiLinfty} for different values of the quantity $n r^2= 4$ ($\full$), 3 ($\broken$), 2 ($\dotted$), 1 ($\chain$), 0.5 ($\longbroken$). }\label{fig:rhovspsi}
\end{figure}

\Fref{fig:rhovspsi} presents the PDFs~\eref{eq:fPsiLinfty} for different values of the quantity $n r^2$. When $n r^2=4$,  all arcs are equal to or less than $\pi/8$ occur with a probability $0.998$. When $n r^2 =10$, this probability is equal to 0.9999998. Since, for $\alpha =\pi/8$, using $\alpha$ instead of $\sin\alpha$ produces an error of less than 3\%, this value can be used as the  threshold for small angles.

Now, let us consider the behavior of~\eref{eq:fPsiLinfty}  when $n r^2 \gg 1$. In this case, almost all the arcs are small, i.e., $\psi \approx 0$, and the number of isolated circles is negligible. Hence, we can omit  the term with the $\delta$-function and ignore the bottom line in~\eref{eq:fPsiLinfty}. Since $\sin \psi \approx \psi$ and $\cos\psi \approx 1$,
the PDF is equal to
\begin{equation}\label{eq:rhoPsi-infty}
f^\infty_{\Psi}(\psi, n )
\approx
4 n r^2\exp(- 4\psi n r^2), \quad n r^2 \gg 1.
\end{equation}
Note, that~\eref{eq:rhoPsi-infty} can be derived in another way, directly from the consideration of a dense system and averaged quantities by omitting consideration of the common case (see \ref{sec:appendix}).

$\eta = na$, where $a$ is the area of the deposited object, is known as the filling factor (see, e.g.,~\cite{Mertens2012PRE}). $\phi = 1 - \mathrm{e}^{-\eta}$ is the surface coverage. In the case of discs with radii $r$, $\eta = \pi r^2 n$, hence,
$$
f^\infty_{\Psi}(\psi, \eta )
\approx
\frac{4\eta}{\pi}\exp\left(- \frac{4\eta\psi}{\pi} \right), \quad \eta \gg 1.
$$

\section{Application to transparent electrodes}\label{sec:application}

Recently, a formula for the sheet resistance of dense homogeneous random resistor networks (RRNs)
has been proposed~\cite{Kumar2016JAP,Kumar2017JAP}. The derivation of this formula is based on the assumption that, along such a system, the electrical potential drops linearly. This idea has been adapted to ring-based conductive films~\cite{Azani2019JAP}. The following summary of the idea basically follows this article~\cite{Azani2019JAP}.

\begin{figure}[!htbp]
  \centering
  \includegraphics[width=0.5\columnwidth]{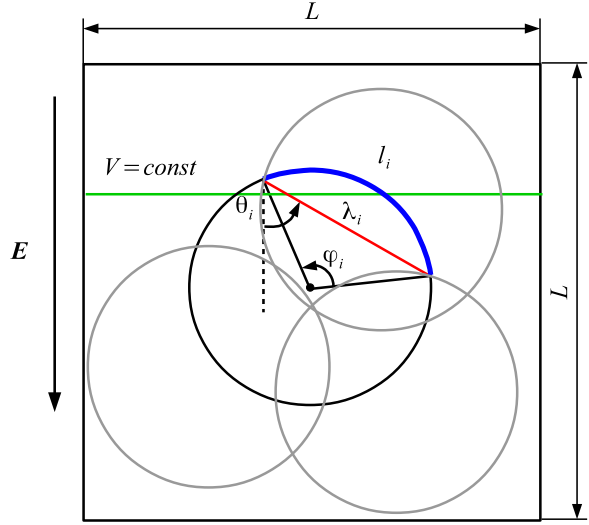}
  \caption{Diagram showing the derivation of the sheet resistance. To provide a clearer view, the rings are exaggerated. The arc between the two intersection points has length $l_i$ and subtends the central angle $\varphi_i$. The corresponding chord has the length $\lambda_i$. $\theta_i$ is the angle between this chord and the vertical.
  \label{fig:Kumar}}
\end{figure}

Let us consider an insulating film of size $ L \times L $. Equally-sized conductive rings $r$ are randomly deposited onto this film. Since the system is supposed to be dense, almost all the rings belong to the giant component, i.e., they are involved in the electrical conductivity. A potential difference, $V$, is applied to the opposite borders of the film (\fref{fig:Kumar}). Due to the  linear potential drop along the system, the equipotential (isopotential) is a straight line.

Potential difference between two intersection points (junctions), which are the ends of the arc (\fref{fig:Kumar}), is proportional to the length of the vertical projection of the chord.
$$
\Delta V_i = \frac{V \lambda_i \cos \theta_i}{L}.
$$
The electrical conductivity of an arc is inversely proportional to its length, $l_i$,
$$
\sigma_i = \frac{\sigma}{l_i},
$$
where $\sigma$ is the conductivity of the wire.
The electrical current through this arc is
$$
I_i = \frac{\sigma_i V \lambda_i \cos \theta_i}{L} = \frac{\sigma V  \lambda_i \cos \theta_i}{ L l_i}.
$$
The total electrical current through the film is equal to
$$
I = \frac{\sigma V }{L} \sum_{i=1}^{N_\mathrm{V}} \frac{\lambda_i}{l_i} \cos \theta_i,
$$
where the summation extends over the all arcs intersecting an equipotential.
The film resistance is
$$
R = \frac{L}{\sigma \sum_{i=1}^{N_\mathrm{V}} \frac{\lambda_i}{l_i} \cos \theta_i}.
$$
Since we are considering a square sample, this quantity is the same as the sheet resistance.
All orientations of the arcs are assumed to be equiprobable.

The sample mean
$$
\frac{1}{N_\mathrm{V}} \sum_{i=1}^{N_\mathrm{V}} \frac{\lambda_i}{l_i} \cos \theta_i
$$
is supposed to be close to the expected value
$$
\frac{1}{\pi}\int_{-\pi/2}^{\pi/2} \int_{0}^{2\pi} \frac{\lambda(\varphi)}{l(\varphi)} f(\varphi,n) \, \rmd\varphi \cos \theta \, \rmd\theta,
$$
where $f(\varphi,n) $ is the PDF of the arc size, $\varphi$, when the number density of the rings is equal to $n$.
Moving from summation to integration, we have
$$
\frac{1}{N_\mathrm{V}}  \sum_{i=1}^{N_\mathrm{V}}  \frac{\lambda_i}{l_i} \cos \theta_i \to
\frac{1}{\pi}\int_{-\pi/2}^{\pi/2} \int_{0}^{2\pi} \frac{\lambda(\varphi)}{l(\varphi)} f(\varphi,n) \, \rmd\varphi \cos \theta \, \rmd\theta = \frac{2}{\pi} \int_{0}^{2\pi} \frac{\lambda(\varphi)}{l(\varphi)} f(\varphi,n) \, \rmd\varphi,
$$
where $\varphi$ is the central angle, which is subtending by this arc. Since arc length is equal to
$l(\varphi) = r \varphi$, while the chord length is $\lambda(\varphi) = 2 r \sin \frac{\varphi}{2},$
\begin{equation}\label{eq:chord-to-arc}
 Z(n) = \int_{0}^{2\pi} \frac{\lambda(\varphi)}{l(\varphi)}f(\varphi,n)  \, \rmd\varphi = \int_{0}^{2\pi} \frac{2 \sin \frac{\varphi}{2}}{\varphi} f(\varphi,n)  \, \rmd\varphi.
\end{equation}
When the distance between the ring center and an equipotential (\fref{fig:Kumar}) does not exceed $r$, the equipotential intersects this arc. On average, each equipotential intersects each of $2 r n L$ rings twice.

Thus, in order to estimate the electrical conductance of the sample, the integral~\eref{eq:chord-to-arc} should be calculated and the number of intersections of each  equipotential with different arcs, $N_\mathrm{V}$, should be found.

\begin{figure}[!hbpt]
  \centering
  \includegraphics[width=0.6\textwidth]{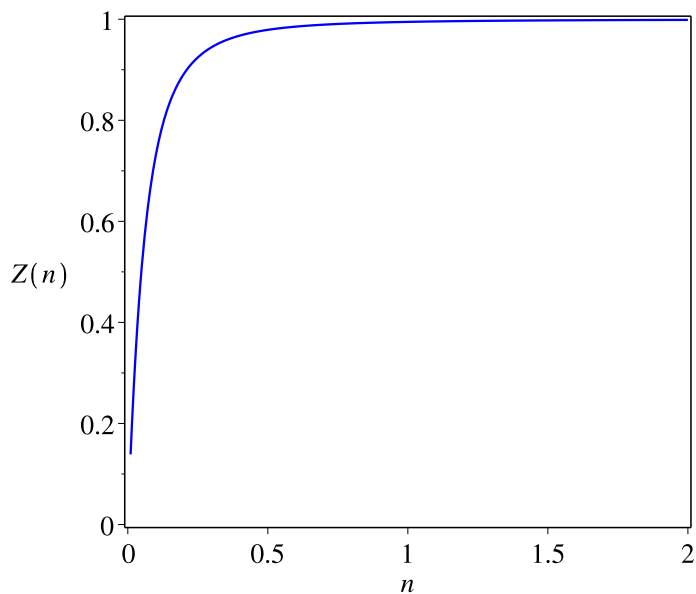}
  \caption{Expected value of the ratio of the chord length to the arc length~\eref{eq:chord-to-arc} against the number density $n$. The plot was obtained by numerical integration of~\eref{eq:chord-to-arc} using the PDF from~\eref{eq:fPsiLinfty}.}\label{fig:arc-chord}
\end{figure}

Using the PDF for large and dense systems~\eref{eq:fpsi},
$$
Z(n) = \int_{0}^{\infty}  \frac{2 \sin \frac{\varphi}{2}}{\varphi} 4 n r^2\exp(- 4\varphi n r^2)  \, \rmd\varphi = 8 n r^2 \arctan \frac{1}{8 n r^2}= 8 n r^2 \cot^{-1} \left( 8 n r^2 \right).
$$
Since the PDF decreases rapidly, we have changed the upper limit of the integral ($2\pi \to \infty$). The plot of this PDF is indistinguishable from the plot of the PDF from~\eref{eq:fPsiLinfty} even for $n \approx 0$ (\fref{fig:arc-chord}).

\begin{figure}[!hbtp]
  \centering
  \includegraphics[width=0.7\textwidth]{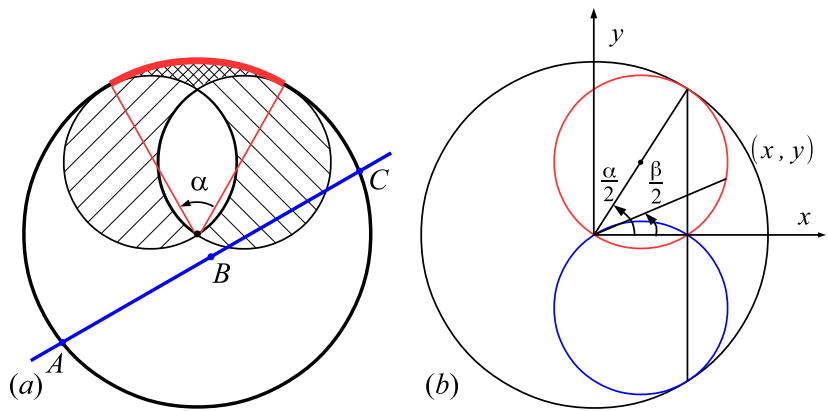}\\
  \caption{Chord $AC$ intersects the arc subtending the central angle $\alpha$ twice, when the middle  of the chord (point $B$) is located within the double hatched area; this chord intersects the arc  only once, when point $B$ is located within the hatched area.}\label{fig:doubleintersection}
\end{figure}

Now, we turn to the number of different arcs intersected by an equipotential,  $N_\mathrm{V}$.
Consider an arbitrary straight line intersecting an arc of radius $r$. Let the intersection points be $A$ and $C$. The middle of the chord $AC$ is denoted as $B$ (\fref{fig:doubleintersection}($a$)). Point  $B$ uniquely defines a straight line, except for the case when $B$ is the center of this circle. When point $B$ is located within the double hatched area, the chord twice intersects the arc  subtending the central angle $\alpha$. When point $B$ is located within the hatched area, the chord intersects this arc only once. When point $B$  is located within the empty area, the chord cannot intersect this arc.

Let the distance between point $B$ and the circle center be $x$. The probability of finding point $B$ within the double hatched region is equal to the ratio of the length of an arc with radius $x$, completely located within the double hatched region ($\beta x$), to the total circumference ($2 \pi x$)
$$
\frac{\beta}{2 \pi}.
$$
Let the origin of a coordinate system coincide with the circle center, while the abscissa goes through the ``beak'' of the double hatched area (\fref{fig:doubleintersection}($b$)). Then, the coordinates of the ``beak'' are
$$
\left(r\cos\frac{\alpha}{2},0\right).
$$
The polar equations of the circles of radii $r/2$ that bound the double hatched region, in the polar coordinate system $(\rho,\phi)$ are
$$
\rho = 2 ( x_0 \cos  \phi \pm y_0 \sin \phi), \quad x_0 = \frac{r}{2}\cos\frac{\alpha}{2}, \quad y_0 = \frac{r}{2}\sin\frac{\alpha}{2}
$$
or
$$
\rho = r \cos\left( \frac{\alpha}{2} \mp \phi\right).
$$
Thus,
$$
\phi(\rho) = \pm \left( \frac{\alpha}{2} - \arccos \frac{\rho}{r}  \right).
$$
The probability, that the chord twice intersects the arc, is
$$
\int_{r\cos(\alpha/2)}^{r} \frac{\phi(\rho)}{\pi} \, \rmd\rho =
\frac{1}{\pi}\int_{r\cos(\alpha/2)}^{r} \left( \frac{\alpha}{2} - \arccos \frac{\rho}{r}  \right) \, \rmd\rho = \frac{r}{\pi} \left( \frac{\alpha}{2} - \sin\frac{\alpha}{2} \right).
$$
The probability, that the chord intersects the arc only once, is
$$
\frac{1}{\pi}\left(\int_{r\cos(\alpha/2)}^{r} 2 \arccos \frac{\rho}{r}  \, \rmd\rho  + \int_{0}^{r\cos(\alpha/2)} \alpha \, \rmd\rho \right)= \frac{2r}{\pi} \sin\frac{\alpha}{2}.
$$
Finally, the probability, that a straight line that intersects an arc, intersects this arc exactly twice is equal to the ratio of the probability, that this straight line intersects the arc twice, to the probability, that this line intersects the arc any number of times,
$$
\frac{\frac{\alpha}{2} - \sin\frac{\alpha}{2} }{\frac{\alpha}{2} + \sin\frac{\alpha}{2} } \approx \frac{\alpha^2}{48}.
$$
Thus, the fraction of arcs that are twice intersected by an equipotential,
$$
\int_0^\infty \frac{\alpha^2}{48} 4 n r^2\exp(- 4\alpha n r^2) \, \rmd\alpha = \frac{1}{384 n^2 r^4},
$$
is very small, i.e., with high precision, the number of different arcs intersected by an equipotential is equal to double the number of intersected circles, i.e.,
$$
N_\mathrm{V} = 4 r n L.
$$
In such a way, for large values of the number density, $n$, the sheet resistance of the ring-based conductive film is equal to
$$
R = \frac{\pi}{8 r n \sigma }.
$$
This is exactly the same formula that was obtained previously using numerical estimations of arc length distributions~\cite{Azani2019JAP}.

\section{Conclusion}\label{sec:conclusion}
We considered the distribution of arc sizes in a system of equal-sized rings, $r$, the centers of which are i.i.d. within a square domain of size $L \times L$ with PBCs ($L > 2r$). For an arbitrary number of rings, $N$, and domain size, $L$, we derived the PDF~\eref{eq:fpsi}. When $L \to \infty$, the PDF reduces to~\eref{eq:fPsiLinfty}. When the number density is large ($n \gg 1$), additional simplification is possible, which leads to~\eref{eq:rhoPsi-infty}. The latter PDF was used to estimate the sheet resistance of ring-based conductive films. This estimation was based on the assumption that the potential drop along a homogeneous dense random system is expected to be linear~\cite{Kumar2016JAP,Kumar2017JAP,Azani2019JAP}. \Eref{eq:rhoPsi-infty} allowed us to draw the  conclusion that, in such the systems, the ratio of the arc length to the chord length is close to unity since almost all the arcs are short. Moreover, the probability, that a straight line intersects the same arc twice, is negligible for the same reason.

The consideration of rings with a given distribution of radius sizes looks like a promising future direction since, in real world systems, size dispersity of rings is observed~\cite{Azani2019JAP}.

\appendix
\section{PDF of large and dense systems}\label{sec:appendix}
Expected number of intersections of a circle with other circles is
$4 \pi n r^2$, hence, any circle in average is divided into $ 8 \pi n r^2$ arcs.
Expected value of the central angle subtended by the arc is
\begin{equation}\label{eq:meanphi}
\langle \varphi \rangle = \frac{2\pi}{ 8 \pi n r^2} \approx \frac{1}{4 n r^2}.
\end{equation}
The number of arcs per unit area is $n_\mathrm{a} = 8 \pi n^2 r^2$.
Note that
\begin{equation}\label{eq:nplusdn}
8 \pi (n+dn)^2 r^2 = 8 \pi r^2  n^2  \left( 1 + \frac{dn}{n}\right)^2 \approx 8 \pi r^2  n^2  \left( 1 + 2\frac{dn}{n}\right)
\end{equation}
and
\begin{equation}\label{eq:fnplusdn}
 f_\Psi(\varphi,n + dn) =  f_\Psi(\varphi,n) + \frac{\partial f_\Psi(\varphi,n)}{\partial n} dn.
\end{equation}

Let the random variable  $\Omega$ be the central angle corresponding to an arc at which a point, randomly thrown  on a random circle, falls. Its PDF is
$$
f_{\Omega}(\omega, n) = \frac{\omega f_{\Psi}(\omega, n)}{\int_{0}^{2\pi}\psi f_{\Psi}(\psi, n) \, d \psi},\; (0\leqslant\omega\leqslant 2\pi).
$$
Expected angle
$$
 \langle\varphi \rangle = \int_{0}^{2\pi}\psi f_{\Psi}(\psi, n) \, d \psi
$$
can be calculated without a knowledge of the explicit kind of the PDF (see~\eref{eq:meanphi}). Hence,
$$
f_{\Omega}(\omega, n) = 4 n r^2\omega f_{\Psi}(\omega, n).
$$

We define arcs corresponding to central angles  $(\varphi,\varphi+d\varphi)$ as $\varphi$-arcs. Let $\#(n)$ be the average number of $\varphi$-rcs per unit area. We define all arcs belonging to initially deposited circles as \emph{old} arcs. Let $dn$ rings per unit area are added to the  $\mathcal{D}$. Note that
$$
\#(n)  =  8 \pi n^2 r^2 f_\Psi(\varphi,n) d\varphi,
$$
$$
 \#(n + dn )  =  8 \pi (n + dn)^2 r^2 f_\Psi(\varphi,n + dn) d\varphi.
$$
Accounting~\eref{eq:nplusdn} and \eref{eq:fnplusdn},
$$
 \#(n + dn ) =  8 \pi n^2 r^2 \left( f_\Psi(\varphi,n) + \frac{2 f_\Psi(\varphi,n)}{n} dn  + \frac{\partial f_\Psi(\varphi,n)}{\partial n} dn  \right)  d\varphi .
$$
Variation of the number of $\varphi$-arcs per unit area when the number of arcs per unit area changes by value $dn$ is
\begin{equation}\label{eq:lhs}
 \#(n + dn ) - \#(n)  = 8 \pi n^2 r^2 \left(\frac{2 f_\Psi(\varphi,n)}{n}  + \frac{\partial f_\Psi(\varphi,n)}{\partial n} \right) dn \,  d\varphi.
\end{equation}

Let
\begin{itemize}
  \item $\Delta_1$ be the average number of $\varphi$-arcs per unit area belonging to newly added circles.
  \item $\Delta_2$ be the average number of  $\varphi$-arcs per unit area arising due to intersections of old arcs with newly added circles.
  \item $\Delta_3$ be the average number of $\varphi$-arcs per unit area destroyed due to intersections of old arcs with newly added circles.
\end{itemize}

Obviously, that the equality
\begin{equation}\label{eq-main}
\#(n + dn) - \#(n)
= \Delta_1 + \Delta_2 - \Delta_3
\end{equation}
is valid.

\begin{equation}\label{eq:Delta1}
\Delta_1 = \frac{\#( n + dn )}{ (n + dn) } dn = 8 \pi (n + dn) r^2  f_\Psi(\varphi, n + dn )  d\varphi \, dn.
\end{equation}
\begin{equation}\label{eq:Delta2}
\Delta_2 = 16 \pi (n + dn) r^2  f_\Psi(\varphi, n + dn ) d\varphi \, dn.
\end{equation}
Therefore,
\begin{equation*}
\eqalign{
\fl
\Delta_1 + \Delta_2 = 3 \cdot  8 \pi (n + dn) r^2 f_\Psi(\varphi, n + dn )  d\varphi \, dn  =\\= 3 \cdot  8 \pi n r^2 \left( 1 + \frac{dn}{n}\right) \left( f_\Psi(\varphi, n ) + \frac{\partial f_\Psi(\varphi,n)}{\partial n} dn \right) d\varphi \, dn.
}
\end{equation*}
Omitting terms with $(dn)^2$,
$$
\Delta_1 + \Delta_2 = 3 \cdot  8 \pi n r^2 f_\Psi(\varphi, n )  d\varphi \, dn.
$$

\begin{equation}\label{eq:Delta3}
\Delta_3 = 8 \pi n r^2 \cdot  4 n r^2\varphi f_{\Psi}(\varphi, n) d\varphi \, dn.
\end{equation}

Substituting~\eref{eq:lhs}, \eref{eq:Delta1}, \eref{eq:Delta2}, and \eref{eq:Delta3} into~\eref{eq-main},
\begin{equation*}
\eqalign{
\fl
8 \pi n^2 r^2 \left(\frac{2 f_\Psi(\varphi,n)}{n}  + \frac{\partial f_\Psi(\varphi,n)}{\partial n} \right) dn \,  d\varphi =\\=
3 \cdot  8 \pi n r^2 f_\Psi(\varphi, n )  d\varphi \, dn
 -
8 \pi n r^2 \cdot  4 n r^2\varphi f_{\Psi}(\varphi, n) d\varphi \, dn.
}
\end{equation*}
Dividing by $8 \pi n^2 r^2 d\varphi \, dn $,
\begin{equation*}
\frac{2 f_\Psi(\varphi,n)}{n}  + \frac{\partial f_\Psi(\varphi,n)}{\partial n} =
\frac{3 f_\Psi(\varphi, n )}{n}  - 4 r^2\varphi f_{\Psi}(\varphi, n).
\end{equation*}
Finally,
\begin{equation}\label{eq:ODEf}
\frac{\partial}{\partial n}f_{\Psi}(\varphi, n)
=\left(  \frac{ 1}{n} -  4 r^2 \varphi \right)   f_{\Psi}(\varphi, n).
\end{equation}
In the solution of~\eref{eq:ODEf}
$$
  f_\Phi(\varphi,n) = A n \exp(-4nr^2\varphi),
$$
the constant $A$ depends on a way of normalization. Since, in dense systems, the fraction of long arcs is negligible,
$$
1= \int_{0}^{2\pi} f_\Phi(\varphi,n) \, d\varphi \approx \int_{0}^{\infty} f_\Phi(\varphi,n) \, d\varphi,
$$
then $A= 4 r^2$,  In this case, the PDF is
\begin{equation}\label{eq:f}
  f_\Phi(\varphi,n) = 4 n r^2 \exp(-4nr^2\varphi)
\end{equation}
or
\begin{equation}\label{eq:f0}
  f_\Phi(\varphi;\langle\varphi\rangle) =  \frac{1}{\langle\varphi\rangle}\exp\left(-\frac{\varphi}{\langle\varphi\rangle}\right),
\end{equation}
i.e., $\Phi \sim \mathrm{Exp}(\langle\varphi\rangle)$.

\ackn
We would like to acknowledge funding from the Ministry of Science and Higher Education of the Russian Federation, Project No.~3.959.2017/4.6.

\section*{References}

\bibliographystyle{iopart-num}
\bibliography{rings}

\end{document}